\documentclass[11pt,twoside]{article}  
\usepackage{apn3conf}

\usepackage[]{natbib}

\newcommand{\nat}{Nature}

\begin{document}

\title{The Last Hurrah: PPN Formation by a Magnetic Explosion}

\author{Sean Matt\altaffilmark{1}}
\affil{Physics \& Astronomy Department, McMaster University, 
Hamilton ON, Canada L8S 4M1}

\altaffiltext{1}{CITA National Fellow}

\author{Adam Frank and Eric Blackman}
\affil{Physics \& Astronomy Department, University of Rochester,
Rochester NY, U.S.A.\ 14627}

\contact{Sean Matt}
\email{matt@physics.mcmaster.ca}

\paindex{Matt, S.}
\aindex{Frank, A.}
\aindex{Blackman, E.}

\authormark{Matt, Frank, \& Blackman}



\begin{abstract}
We discuss a mechanism by which a giant star can expel its envelope in
an outburst, leaving its core exposed.  The outburst is powered by
rotational kinetic energy of the core, transferred to the envelope via
the twisting of magnetic fields.  We show that, if the core is
magnetized, and if it has sufficient angular momentum, this mechanism
may be triggered at the end of the asymptotic giant branch phase, and
drive a proto-planetary nebula (pPN) outflow.  This explosion of
magnetic energy self-consistently explains some of the asymmetries and
dynamics of pPNe.
\end{abstract}

\section{Introduction \label{sec_intro}}

The formation and shaping of planetary nebulae (PNe) has been
described by a generalized interacting stellar wind (GISW) model
\citep{kwok3ea78, balick87}, in which the central star of a PN
produces a fast wind that interacts with the slower moving, previously
ejected, asymptotic giant branch (AGB) wind.  According to the GISW
model, both the slow and fast winds are driven by radiation, and
asymmetric PN shapes result from asymmetries in the slow, AGB wind,
from magnetic effects in the shocked fast wind \citep[e.g.,][see also
Garc\'ia-Segura in these proceedings]{chevalierluo94,
garciaseguraea99}, or from some combination of the two.

However, some recent observations of proto-planetary nebulae (pPNe,
precursors of PNe) call for a revolutionary change in our
understanding of the driving and shaping of post-AGB outflows.  These
observations are as follows:
\begin{enumerate}

\item{The material ejected during the pPN phase is highly structured,
often containing multiple apparent ejection axes and occasionally
appearing quite chaotic \citep[e.g.,][]{balickfrank02}.  Several pPNe
(e.g., CRL 2688, OH231.8+4.2, IRAS 17150-3224, and IRAS 17441-2411)
exhibit quadrupolar symmetry, consisting of outflowing disks---dense
enough to obscure central starlight---and bipolar lobes or jets.}

\item{Spherical shells of reflected starlight often surrounding highly
structured PNe and pPNe \citep[e.g.,][see also Su in these
proceedings]{terzianhajian00, balick3ea01}.  This suggests that the
AGB wind is spherically symmetric, and so it cannot be responsible for
the asymmetry in subsequent flows.  Thus, pPN outflows must be
self-shaped.}

\item{The momentum carried in pPN outflows is often a few orders of
magnitude larger than can be explained by radiation pressure driving
\citep[][see also Bujarrabal in these proceedings]{knapp86,
bujarrabalea01}, so a new and more powerful wind driving mechanism is
needed.}

\item{The pPN phase is extremely short lived, typically only a few
hundred years.  Yet the observed flows often follow a ``Hubble law''
expansion \citep[$v \propto R$, e.g.,][]{alcoleaea01}, suggesting that
ejecta accelerate in a short time compared to the pPN lifetime, and
afterward follow ballistic trajectories.  Therefore, the flows are
really outbursts, transient winds where the momentum and kinetic
energy suddenly and temporarily increase.}



\end{enumerate}
Together, these key observations seem to require a new ejection
mechanism that is quasi-explosive, is triggered at the end of the AGB
phase, and produces intrinsically asymmetric flow structures.


Here we show that the rotational kinetic energy of the core of an AGB
star can be converted to linear kinetic energy and transferred to the
envelope.  The conversion and transferral of energy happens via the
twisting of magnetic fields anchored in the core, and naturally
results in a complete and asymmetric ejection of the envelope.  The
acceleration of the envelope occurs in a very short time compared to
the typical age of pPNe.  We will describe the conditions necessary
for such an ejection to take place and discuss how stars may achieve
those conditions near the end of the AGB phase.  Our model is general
and may be relevant to other classes of objects, such as SNe, GRB's,
and $\eta$ Carinae.

\section{The Magnetic Explosion \label{sec_explosion}}

Consider an AGB star with a slowly contracting core and an expanding,
convective envelope.  If the star is rotating at all, angular momentum
conservation requires that there will be a region of differential
rotation at the interface between the core and the envelope.
\citet{blackmanea01} have shown that the differential rotation in an
AGB star can drive a solar-like magnetic dynamo, in which magnetic
fields are amplified to an (time-averaged) equilibrium value of
$10^4$--$10^6$ Gauss at the interface.  The magnetic field is mixed by
convection throughout the envelope.  Whether a binary interaction is
required to maintain (or trigger) differential rotation in AGB stars
remains an open question.

Large mass loss rates from the envelope, during the late stages of the
AGB phase, result in a slow expansion of the envelope and a decrease
in its density at all depths.  As the weight of the overlying envelope
decreases, the relative importance of the magnetic field increases.
In particular, there may be a threshold at which the magnetic (plus
thermal) energy within the envelope exceeds its gravitational
potential energy.  At that point, the envelope as a whole would be
driven off in a single, short-lived event.

Due to the extreme difficulty of modeling, self-consistently, the
complex interface dynamo, together with a proper treatment of the
turbulent envelope and mass loss from the top of the convection zone,
we consider here a slightly different and simplified situation.
Rather than having an envelope with a constant magnetic energy and a
decreasing density, we consider a constant density envelope in which
the magnetic energy increases.  Due to our simplifications, the
magnetic energy is generated in a short time and only at the interface
between the core and envelope, where strong differential rotation
twists up pre-existing poloidal magnetic fields.  Though the magnetic
energy is produced only at the base of the envelope (and not mixed
everywhere throughout), we expect the same qualitative behavior
described above, where the magnetic energy exceeds the gravitational
potential energy, and the envelope is driven off in an outburst.

For this initial study, we have carried out time-dependent, numerical
magnetohydrodynamic (MHD) simulations.  We use the simulation code of
\citet{matt02}, which employs a finite difference scheme to solve the
standard, adiabatic ($\gamma = 5/3$), ideal MHD equations, plus source
terms for gravity, on an Eulerian, cylindrical ($r$, $z$), nested
grid.  The code assumes axisymmetry (where $\delta/\delta\phi$ of all
quantities $ = 0$).  We use ``outflow'' boundary conditions (BCs) on
the outermost boundary (though material doesn't leave the grid by the
end of the simulations), and assume reflection symmetry about the
equator ($z = 0$).  The simulation domain, then, consists of an
$r$--$z$ slice through a single quadrant, with the stellar core
specified by a circular, inner BC.

For simplicity, we want to ignore effects of a wind from the core, so
we have chosen a ``solid'' BC such that no matter can flow from or
onto the core.  The core behaves as a perfect electric conductor and
initially is rotating as a solid body and is threaded by a
rotation-axis aligned dipole magnetic field.  The computational domain
(existing between the core boundary and the outer boundary) is
initialized with a hydrostatic envelope whose density falls off as
$R^{-2}$ (where $R$ is the spherical radius).  The envelope is
initially stagnant (no rotation or motion) and is threaded everywhere
by the dipole magnetic field that is anchored to the core.  We also
assume that the mass of the core is much greater than that of the
envelope, so that the envelope's self-gravity can be ignored.

\begin{figure}[ht]
\epsscale{0.9}
\plotone{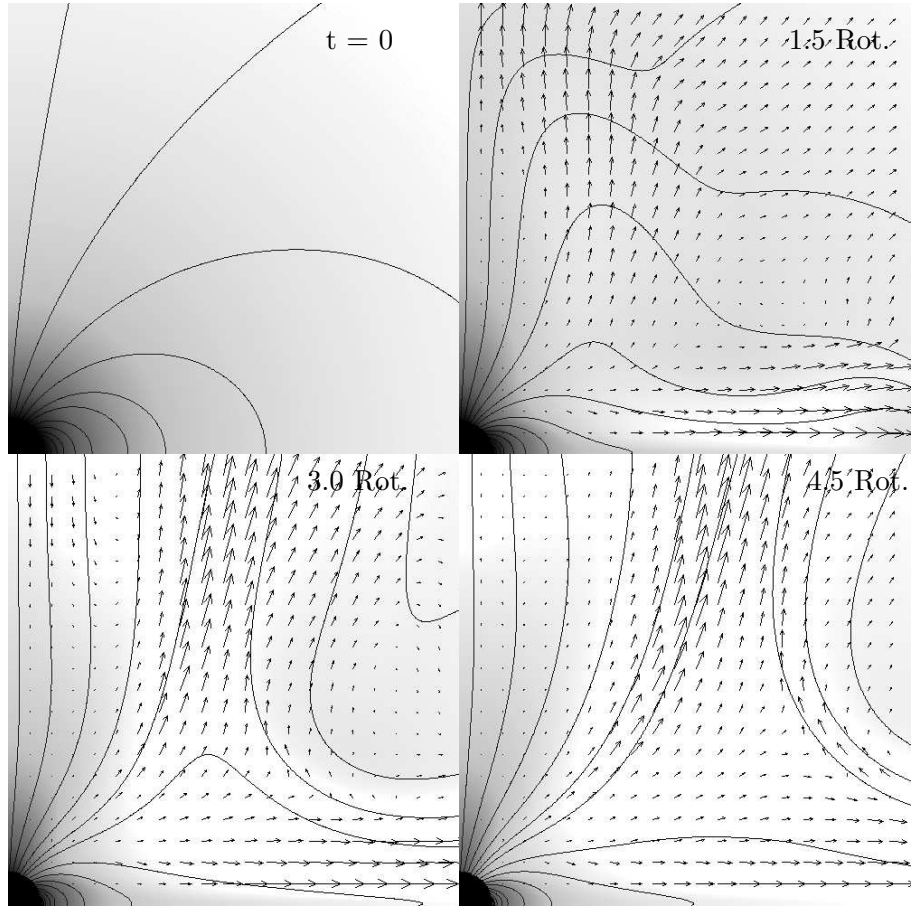}
\caption{Greyscale images of log density (black is highest density),
poloidal field lines, and velocity vectors show the evolution of the
system in the region near the core, spanning $0 \le r$, $z \le 13
R_{\rm core}$.  The core is in the lower left of each panel, and the
rotation axis is vertical.
\label{fig_closemovie}}

\vspace*{-13.9cm}
\hspace*{4cm}
t = 0
\hspace*{5cm}
1.5 Rot.
\vspace*{13.55cm}

\vspace*{-8.1cm}
\hspace*{4cm}
3.0 Rot.
\hspace*{5cm}
4.5 Rot.
\vspace*{7.75cm}

\end{figure}

The upper left panel of figure \ref{fig_closemovie} shows the initial
configuration in the inner part of the computational domain (from the
center, out to 13 core radii, $R_{\rm core}$).  The initial
configuration is static and in equilibrium.  At $t=0$ the core begins
to rotate.  Differential rotation between the core and stagnant
envelope leads to an azimuthal twisting of the dipole field lines just
above the core.  The azimuthal magnetic field $B_\phi$ increases
linearly in time, until it becomes strong enough to be dynamically
important.  At that time, magnetic pressure forces (i.e., $-\nabla
B_\phi^2$), begin to drive material outward, away from the core.

In our chosen configuration, the key parameters are the relative
values of the characteristic speeds in the system---the equatorial
rotation speed of the core ($v_{\rm rot}$), the initial Alfv\'en speed
at the core equator ($v_{\rm A}$), and the escape speed from the
surface of the core ($v_{\rm esc}$)---which are related to the
relative values of various energies in the system.  Note that in
hydrostatic equilibrium, the sound speed is proportional to the escape
speed, so the sound speed (or thermal energy) is not a key,
independent parameter.  Here, we present a case with $v_{\rm
rot}/v_{\rm esc} = 0.1$ (14\% of breakup rotation) and $v_{\rm
A}/v_{\rm esc} = 1.0$.

The upper right, lower left, and lower right panels of figure
\ref{fig_closemovie} show the evolution after 1.5, 3.0, and 4.5
rotations of the core, respectively.  The expansion of newly generated
azimuthal magnetic field is evident in the figure, as it drives the
envelope material outward.  This also leads to an expansion of the
poloidal magnetic field (lines), though some field lines near the
equator remain closed, where material is forced and held in corotation
with the core.  Due to the coupling of the rotation to the dipole
field, $B_\phi$ has a maximum near the core at mid latitudes.  Thus,
magnetic pressure forces point (from strong to weak field), not only
radially away from the core, but also toward the rotation axis and
toward the magnetic equator.  This produces structures in the outburst
such that the expansion is faster toward the equatorial and polar
directions.  Finally, notice that the density near the core decreases
with time.  This is because that region is drained of material by the
expansion of the magnetic field (our ``solid'' core BC prevents the
region from being replenished by mass flow from the core).  This
process is truly a transient phenomenon.

\begin{figure}[ht]
\epsscale{0.9}
\plotone{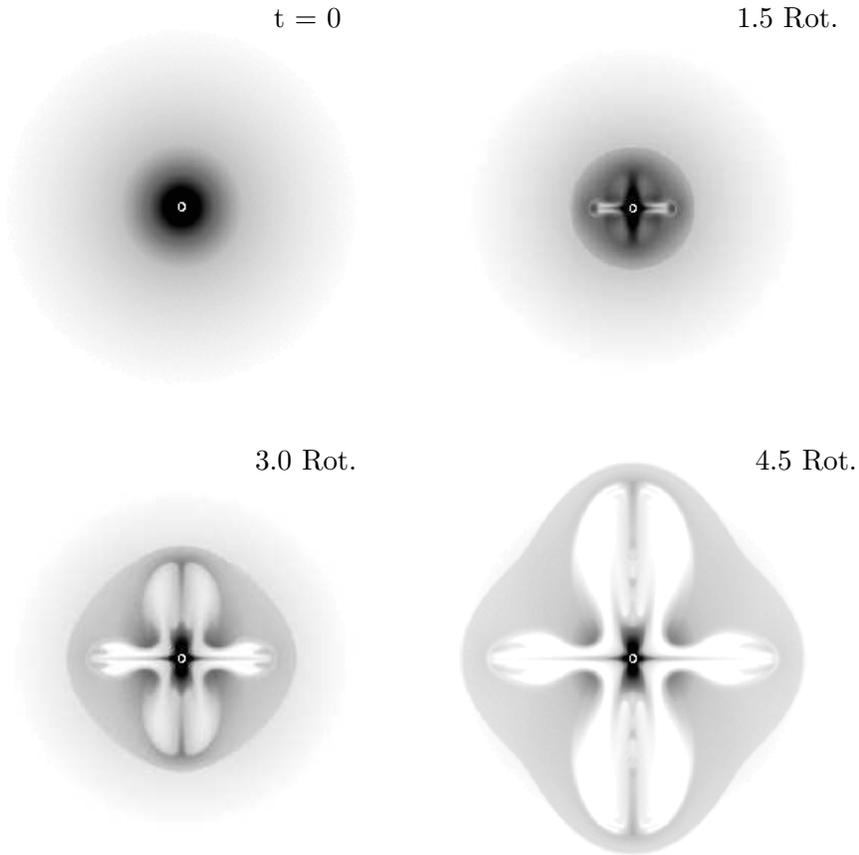}
\caption{Greyscale images of log density (black is highest density)
show the evolution of the system far from the core, spanning
$-130R_{\rm core} \le r$, $z \le 130 R_{\rm core}$.  The core is
indicated by a white circle in the center of each panel, and the
rotation axis is vertical.
\label{fig_farmovie}}

\vspace*{-13.9cm}
\hspace*{4cm}
t = 0
\hspace*{5cm}
1.5 Rot.
\vspace*{13.55cm}

\vspace*{-8.1cm}
\hspace*{4cm}
3.0 Rot.
\hspace*{5cm}
4.5 Rot.
\vspace*{7.75cm}

\end{figure}

The four panels of figure \ref{fig_farmovie} show the evolution of the
system at the same times and for the same simulation as in figure
\ref{fig_closemovie}.  However, the data in figure \ref{fig_farmovie}
is shown on a scale that is 10 times larger than in figure
\ref{fig_closemovie} (by taking advantage of nested simulation grids),
and the data has been reflected about the rotation axis and equator to
better illustrate the flow.

\begin{figure}[ht]
\epsscale{0.5}
\plottwo{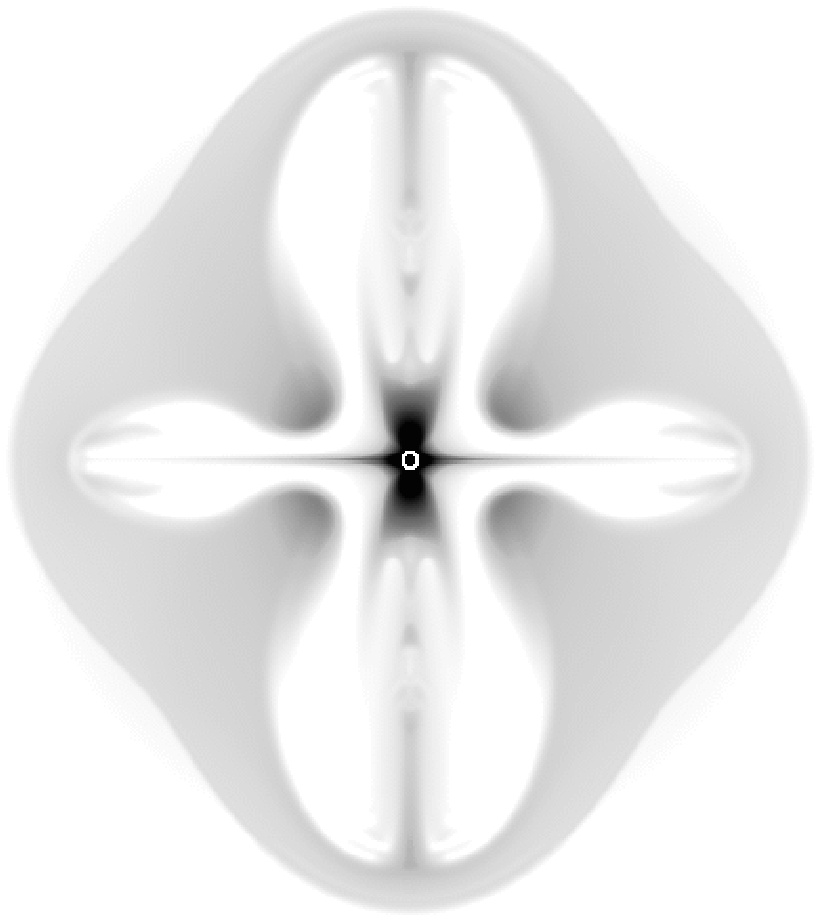}{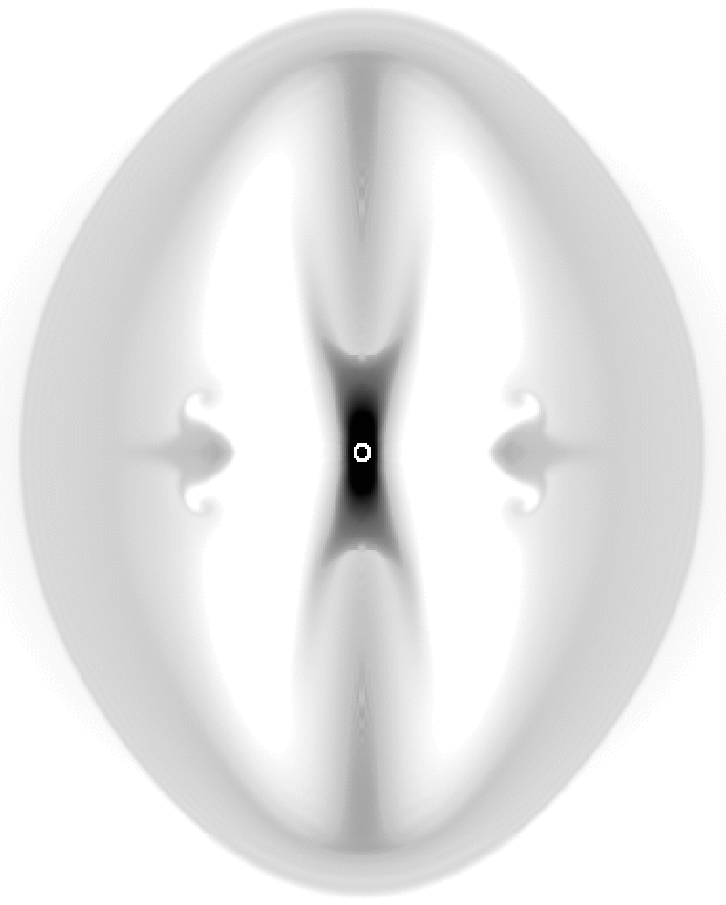}
\caption{Case with initial dipole magnetic field (left) compared to a
case with initial split monopole (right).  The panels have the same
spatial and density scale as figure \ref{fig_farmovie}.
\label{fig_monopole}}
\end{figure}

As the simulation proceeds, one sees a swept up shell of envelope
material moving outward from the core.  In the upper right panel, the
outer edge of this shell is quite spherical, and it represents the
initial pressure wave expanding outward to make room for the expanding
magnetic field.  As the field expansion accelerates and exceeds the
sound speed, it overtakes the spherical pressure wave and begins to
protrude outward along the rotation axis and magnetic equator.  This
is evident in the bottom two panels, where the outer dense region is
swept up envelope material, and the ``hollow'' inner region is
actually filled with magnetic (mostly $B_\phi$) energy.  Note the
strongly quadrupolar shape of this inflating bubble.  Again, the
asymmetry is due to the fact that a spinning dipole produces a
$B_\phi$ that is strongest at mid latitudes, so magnetic pressure
forces direct outward from the core, and also toward the pole and
equator.  The magnetic field responsible for inflating the bubble is
generated by the rotation of the core.  Thus we are seeing the spin
energy of the core being converted into linear kinetic energy of the
envelope (and the core will spin down, as a result).

We have run several simulations with different relative values of
$v_{\rm rot}$, $v_{\rm A}$, and $v_{\rm esc}$.  We find that the
envelope will be fully ejected from the system, as long as the
dimensionless parameter $v_{\rm rot} v_{\rm A} v_{\rm esc}^{-2} \ga
0.1$.  For smaller values, the outburst may occur only at selected
latitudes (e.g., the pole and equator), or, for very small values,
there is no envelope ejection.  For cases where the envelope is
ejected, the expansion speed of the swept up shell is approximately
twice the rotation speed, $v_{\rm rot}$.  Also, various shell shapes
are produced, with different relative speeds in the polar and
equatorial regions.

The enhanced equatorial flow is a feature of the spinning dipole
field.  Different field geometries will not necessarily have an
equatorial enhancement, though there will always be a polar
enhancement.  To demonstrate this more clearly, we ran a simulation
with the same parameters as the data in figures \ref{fig_closemovie}
and \ref{fig_farmovie}, but with a split monopole field geometry
(radial magnetic field, but with a direction reversal at the equator).
The right panel of figure \ref{fig_monopole} shows the split monopole
case after three rotations of the core.  For comparison, the left
panel shows the dipole case (same as the lower right panel of fig.\
\ref{fig_farmovie}).

\section{Discussion}

The magnetic explosion mechanism presented here, if triggered at the
end of the AGB phase, may solve all of the problems discussed in
section 1.  Triggering requires that the field and the rotation of the
core must either be sustained throughout the lifetime of the AGB phase
or be generated near the end of the AGB phase, perhaps stimulated by
binary interaction.  To compare our simulations to real pPNe, we can
scale our parameters to real values.  For example, we can assume a
core mass of $0.5 M_\odot$, a swept up shell mass and size of $0.062
M_\odot$ and $10^4$ AU \citep[for CRL 2688,][see also Kastner in these
proceedings]{bujarrabalea01}, and $v_{\rm esc} = 1000$ km s$^{-1}$.
In that case, the simulation presented in figures \ref{fig_closemovie}
and \ref{fig_farmovie} applies to a core with a radius of $0.2
R_\odot$, $v_{\rm rot} = 100$ km s$^{-1}$, and a dipole field strength
of $2 \times 10^5$ Gauss.  If instead we assume $v_{\rm esc} = 3000$
km s$^{-1}$, the same simulation applies to a core with a radius of
$10^{9}$ cm, $v_{\rm rot} = 300$ km s$^{-1}$, and a dipole field
strength of $10^8$ Gauss.

The time taken to extract all of the rotational kinetic energy from
the core (i.e., the spin-down time) is $\sim 100$ years, comparable to
the typical pPNe lifetime.  This may explain the short acceleration
times apparent in pPNe outflows and the relative lack of quickly
rotating white dwarfs \citep{koesterea98}.

\acknowledgments

We thank the organizers for a productive and enjoyable meeting.  This
research was supported by NSERC, McMaster University, and CITA through
a CITA National Fellowship.





\end{document}